\if 10
  \documentclass{svjour3}                     
  \smartqed  
  \usepackage{graphicx,natbib,epsfig, epstopdf, color,hyperref}
  \journalname{Experimental Economics}
  \title{The Online Laboratory: \\ Using Internet-based Labor Markets for Experimentation} 
  \author{John J. Horton
    \and David G. Rand 
    \and Richard J. Zeckhauser}
  \institute{F. Author \at
              first address \\
              Tel.: +123-45-678910\\
              Fax: +123-45-678910\\
              \email{fauthor@example.com}           
           \and
           S. Author \at
              second address
           }

  \date{Received: date / Accepted: date}

\else

  \documentclass[12pt]{article} 
  \usepackage{longtable,amsthm,tabularx,
    longtable, booktabs, setspace,varioref,
    ifthen,natbib,amssymb,verbatim,dcolumn, graphicx,lscape,setspace,
    amsmath,color,subfigure,threeparttable, hyperref}

  \title{The Online Laboratory: \\ Conducting Experiments in a Real
    Labor Market\footnote{Thanks to Alex Breinin and Xiaoqi Zhu for
      excellent research assistance. Thanks to Samuel Arbesman, Dana
      Chandler, Anna Dreber, Rezwan Haque, Robin Yerkes Horton, and
      Stephanie Hurder for helpful comments, as well as to
      participants in the Online Experimentation Workshop hosted by
      Harvard's Berkman Center for Internet and Society. This research
      has been supported by the NSF-IGERT program ``Multidisciplinary
      Program in Inequality and Social Policy'' at Harvard University
      (Grant No. 0333403).}}
  \author{John J. Horton\footnote{Harvard University.\href{http://sites.google.com/site/johnjosephhorton}{http://sites.google.com/site/johnjosephhorton}.}  \and David
    G. Rand\footnote{Harvard
      University.\href{http://www.people.fas.harvard.edu/~drand}{http://www.people.fas.harvard.edu/\~{}drand}.}
    \and Richard J. Zeckhauser\footnote{Harvard
      University.\href{http://www.hks.harvard.edu/fs/rzeckhau}{http://www.hks.harvard.edu/fs/rzeckhau}.}}
  \date{}

\fi

\usepackage{Sweave}
\begin{document}
\maketitle

\setkeys{Gin}{width = .5 \textwidth}

\begin{abstract}
  Online labor markets have great potential as platforms for
  conducting experiments, as they provide immediate access to a large
  and diverse subject pool and allow researchers to conduct randomized
  controlled trials. We argue that online experiments can be just as
  valid---both internally and externally---as laboratory and field
  experiments, while requiring far less money and time to design and
  to conduct. In this paper, we first describe the benefits of
  conducting experiments in online labor markets; we then use one such
  market to replicate three classic experiments and confirm their
  results. We confirm that subjects (1) reverse decisions in response
  to how a decision-problem is framed, (2) have pro-social preferences
  (value payoffs to others positively), and (3) respond to priming by
  altering their choices.  We also conduct a labor supply field
  experiment in which we confirm that workers have upward sloping
  labor supply curves.  In addition to reporting these results, we
  discuss the unique threats to validity in an online setting and
  propose methods for coping with these threats. We also discuss the
  external validity of results from online domains and explain why
  online results can have external validity equal to or even better
  than that of traditional methods, depending on the research
  question. We conclude with our views on the potential role that
  online experiments can play within the social sciences, and then
  recommend software development priorities and best practices.
\end{abstract}

\doublespacing 
  
\section{Introduction}
Some of the first experiments in economics were conducted in the late
1940s to test predictions from the emerging field of game
theory. While the research questions were complex, the tools were
simple; paper, pencils, blackboards and marbles were sufficient instruments
to present stimuli and capture subjects' choices and
actions.\footnote{For an historical perspective on early
  experimentation in economics, see \cite{kagel1995handbook}.} By the
early 1990s, researchers had developed tools for conducting
experiments over local computer networks, with subjects receiving
stimuli and making decisions via computer terminal
\citep{fischbacher2007z}. This development made it easier to carry out
game play and collect data, easing the burden placed on
experimenters. However, the advantages of computer-mediation were not
merely logistical; experimenters also gained greater control over the
flow of information and thereby reduced the relevance of potentially
confounding factors.  Computer-mediation quickly became the primary
means for conducting laboratory experiments.  Nevertheless, even
today, human subjects are still brought into physical labs to
participate in experiments that could, at least in principle, be
conducted over the Internet with less inconvenience to the subjects
and less cost in time and money to the experimenters. Why have we not
built the online laboratory?

The case for online experimentation is strong; the clearest advantage
of the online laboratory is that there is no need to physically
aggregate subjects and compensate them for their travel.  This
advantage is so obvious that social scientists recognized it more than
a decade ago. In 1997, the National Science Foundation sponsored a
workshop called NetLab to investigate the potential of online
experimentation \citep{bainbridge2007scientific}.  That workshop's
report identified the major advantages of online experimentation,
including larger sample sizes, greater subject diversity, and
lengthier experiments.  NetLab's participants optimistically
concluded:

\begin{quote}
If the nascent laboratory experimental approach is encouraged and is
coupled with new technological innovations, then the SBE [social,
behavioral, and economic sciences] disciplines will be primed for major
scientific advances.
\end{quote}

The ``if'' in that conclusion hinged on some rather mundane obstacles:
funding, particularly for software development, and technical
training.  Thirteen years have passed; and yet, despite an explosion
in the size, usage, and capabilities of the Internet, online
experiments are still relatively rare, particularly in economics.
During the same period, both field and traditional laboratory
experiments have become far more common \citep{levitt2009field}.
We believe that the practical problems of (1) recruiting subjects and
paying them securely and (2) assuring internal validity---and not
those of funding or training constraints---have limited the
development of online experimentation.

In this paper, we argue that a recent development, the emergence of
online labor markets, effectively and efficiently addresses both the
recruitment/payment problem and the internal validity problem. Online
labor markets also possess distinct advantages over both laboratory
and traditional field experiments. These markets allow workers from
around the world to perform tasks amenable to remote completion, such
as data entry, computer programming, graphic design and clerical work
\citep{frei2009}.  These markets, although designed for other
purposes, make it possible to recruit large numbers of subjects who
are ready and able to participate in experiments.  The subjects
themselves are much more diverse and much less experiment-savvy than
subjects in traditional laboratories. Even better, online labor
markets are ideal for ``experimenter-as-employer'' experiments, which
overcome some of the artificiality critiques of conventional laboratory
experiments, which have the subjects work at contrived tasks and with full
knowledge of the experimental nature of their ``jobs.''

While greater subject diversity and market context are attractive
features, the prime advantage of online labor markets is that they
permit the level of control needed for causal inference. The creators
of these online labor markets---for their own, unrelated
purposes---have made it easy to make individual-specific payments,
screen out users who do not have valid accounts with the market and
prevent workers/subjects from communicating with each
other. Researchers can use these features to conduct internally valid
experiments online. Furthermore, for certain kinds of research
questions in economics, field experiments conducted in online labor
markets can have a very high level of external validity---a point we
will discuss in depth.

Despite the benefits they offer, online experiments raise issues not
frequently encountered in either the laboratory or the field.  Just as
television shows are not filmed plays, online experiments are not
simply laboratory experiments conducted online. This paper discusses
these differences and pays close attention to the unique challenges of
online experimentation. Despite the caveats and pitfalls, the proof
of the value of online experiments is in our replications; we quickly,
cheaply, and easily reproduce a handful of experimental results
known to have external validity. Although we will discuss
\emph{why} online experiments work, the key point is that they
\emph{do} work, at least in the cases we have selected for replication.

In Section \ref{sec:olm} we elaborate on our claim that online
experiments overcome the classic challenges to causal inference. To
back up these claims, we successfully reproduce a series of classic
experimental results in Section \ref{sec:rep}. While these
confirmations support our argument, they are not sufficient to
establish the validity of online methods.  For this reason, in Section
\ref{sec:intval} we highlight the classic threats to valid causal
inference and discuss how they can be overcome in an online
setting. In Section \ref{sec:extval} we then discuss the external
validity of online experimental results. That examination leads naturally to
Section \ref{sec:designs}, in which we discuss the pros and
cons of different experimental designs, such as the
experimenter-as-employer and traditional laboratory economic
games. In addition to creating exciting opportunities for research,
online experiments also pose some ethical challenges, which we address
in Section \ref{sec:ethics}. We conclude in Section \ref{sec:concl}
with our thoughts on the future of the online laboratory.

\section{Online Labor Markets and Experimentation} \label{sec:olm}
At present, the most useful online labor markets, from an
experimentation standpoint, are ``all-purpose'' labor markets where
buyers contract with individual sellers \citep{horton2010olm}.  Some
of the larger markets in this category include oDesk, Elance, Guru and
Amazon's Mechanical Turk (MTurk).\footnote{There are other online labor
  markets, structured more like tournaments or prize-based contests,
  that are less relevant for experimental purposes.}  Each of these
sites is potentially amenable to experimentation, but MTurk currently
offers the best venue due to its robust application programming
interface (API) and pricing structure.  Even better, experimenters
can tap the deep knowledge and experience of web developers to create
practically any interface or environment imaginable. All of the
experiments discussed in this paper were conducted on MTurk.

Amazon claims that the MTurk workforce is larger than $100,000$, but
provides no details about its exact size or composition.  In
conducting our own experiments, we have had no difficulty in rapidly
recruiting very large samples.  If one were to try to build a site for
online experimentation ``from scratch,'' assuming that the technical
hurdles could be overcome, the biggest challenge---as anyone who has
started a lab knows---would be the familiar chicken-and-egg problem of
attracting users.  Few subjects would register unless many potential experiments were already available, and, absent many subjects,
few experimenters would register their experiments.  Online labor
markets solve the chicken-and-egg problem.

\subsection{The advantages of recruiting from labor markets}
Online labor markets offer the benefits associated with recruiting
subjects from an active labor market.  In addition to allowing for
truly massive samples, online labor markets provide diverse samples of
both high- and low-skilled individuals, from a wide range of
countries. One especially useful dimension of subject diversity is
inexperience with economic games.\footnote{Participants are probably
  less experiment-savvy than subjects recruited online, at large and
  specifically for participation in economic experiments.  They are
  certainly less savvy than typical subjects at university research
  laboratories.}  Further, by using subjects from less-developed
countries, experimenters can create relatively high-stakes games for
far less money than would be needed if using subjects from developed
countries.

Experimenters are not required to tell subjects hired from online labor
markets that they are participating in an experiment. For experiments
that use classic economic games, subjects might guess they are in a
study of some kind; but for real-effort, market-appropriate tasks,
workers are unlikely to suspect the academic nature of their work.
This advantage is enormous, as one of the sharpest critiques of the
experimental method in economics is the inherent artificiality of
randomized controlled trials in which subjects are aware of the
experiment.

The subjects' lack of knowledge rules out experimenter effects in which the subjects try to produce the effect they believe the experimenters
expect. It also rules out ``John Henry'' effects, in which subjects exert great effort
because they treat the experiment game like a competitive
contest.  For economics in particular, the fact that subjects
recruited from online labor markets are already in work environments
is a tremendous advantage, as subjects are already making
consequential economic decisions. As a result, they are likely to view
any task or game from an economic frame of mind.

Experimenter and John Henry effects arise because the subjects know that
the setting is experimental. Related problems arise when the subjects know
about other experimental situations. Subjects in a less desirable treatment might
be upset by their bad luck, which might affect their
behaviors. \cite{cook1979quasi} call this ``demoralization.'' Furthermore, even if subjects remain
unaware of an experiment and of the nature of the treatment, agents of
the experimenter might affect outcomes by, on their own initiative, trying to equalize
outcomes, such as by imitation of other treatments or by compensatory
equalization \citep{cook1979quasi}. In online experiments in which subjects have no knowledge of the treatments received by others, the threat of demoralization is minimal. Additionally, in online settings, the
experimenter requires no agents, so the comparisons
between treatment and control groups yield untainted and more accurate
results.

\subsection{Obtaining control and promoting trust} 
Most of the concerns of online market creators mirror those of
would-be experimenters.  Employers worry that workers with multiple
accounts might place phony bids or manipulate the reputation system by
leaving phony feedback. Similarly, experimenters worry that a subject
with multiple accounts might participate in an experiment multiple
times.  The creators of online labor markets do not want workers to
communicate with each other, as it could lead to collusion.
Experimenters also worry about workers discussing the
details of experiments with each other and possibly colluding.  Finally, both employers and
experimenters need ways to pay individuals precise amounts of money as
rewards for their actions and decisions.

It is now easy to hire and pay workers within the context of online
labor markets, yet still quite difficult to do the same online, but outside
of these markets.  The problem is not technological.  The type and
quality of communication---email, instant messenger services,
voice-over-IP---do not depend on whether the buyer and seller are
working inside or outside the online market, and banks have been
electronically transferring funds for decades.  The problem is that it
is difficult to create trust among strangers.  Trust is an issue not
only for would-be trading partners, but also for would-be
experimenters.  The validity of economics experiments depends heavily upon trust,
particularly subjects' trust that the promulgated rules will be
followed and that all stated facts about payment, the identities of
other subjects, etc., are true.  This need for trust is a good reason to
embed experiments in online labor markets, because the creators of
these markets have already taken a number of steps to foster trust.  These efforts
make both commerce and valid experimentation possible.

All online labor markets use reputation systems to create lasting,
publicly-available reputations---reputations that are sacrificed if
either buyers or workers behave unfairly
\citep{resnick2000reputation}.  The market creators proactively screen
out undesired participants by taking steps, such as requiring a bank account or
valid credit card, before either buyer or seller is allowed to join.
With persons who have been accepted, the market creators actively manage memberships and
suspend bad actors, creating a form of virtuous selection not found in
traditional markets.  One kind of bad actor is the non-human, that is,
automated scripts that fraudulently perform ``work.''  To combat this
potential problem, all sites require would-be members to pass a
CAPTCHA, or ``completely automated public Turing test to tell
computers and humans apart'' \citep{von2003captcha}.\footnote{At least
  on MTurk, there is some danger of malicious users writing scripts that
  automatically accept and complete ``Human Intelligence Tasks,'' or
  HITs.  However, these attempts are trivially easy to detect for
  anything more complicated than a single yes/no question. Further, in
  our experience, jobs that allow workers to only complete one unit of
  work (which is almost always the case with experiments) do not
  attract the attention of scammers writing scripts.} With proper
precautions, it is unlikely that computers would show up as subjects,
or that any worker/subject would believe they were playing against a
computer.

While the ``Turing test'' form of trust is important, the mundane but
perhaps more critical requirement is that workers/subjects trust that
buyers/experimenters will actually follow the rules that they
propose.  To encourage this form of trust, many of the online labor
markets require buyers to place funds in escrow, which prevents buyers
from opportunistically refusing to pay after taking delivery of the
worker's output (which is often an easy-to-steal informational good). 
In many markets, there is some form of dispute arbitration, which
encourages the belief that all parties are acting in the shadows of an
institution that could hold them accountable for their actions,
further promoting trust.  Perhaps unsurprisingly, survey evidence
suggests that workers in MTurk believe that their online bosses are as
fair as employers in their home countries \citep{horton2010condition}.

\subsection{Experiments in MTurk}
Online experiments in MTurk are quite simple to run. An advertisement
is placed for the experiment via the same framework used to advertise
jobs in the online labor market.  Depending on Institutional Review
Board (IRB) requirements and the nature of the experiment, the
advertisement may or may not explicitly state that the task is part of
an academic study.  In MTurk, subjects may begin work on a job without
prior approval from the employer.  To make this possible, the
experimenter needs to create a website that gives the subjects
instructions, records their choices, provides them with information as
the game progresses and determines their payoffs. Websites can
incorporate a range, from simple surveys made with off-the-shelf
software to custom-built, elaborate interfaces with designs limited
only by time and resources. Currently, there are a number of projects
under development to build a set of tools for common experimental
tasks, such as assigning subjects to treatment and control groups.

\section{Experiments in the Online Laboratory} \label{sec:rep}
This section discusses research conducted in the online
laboratory---both our own work and that of others.  For this paper, we
conducted three experiments similar to those normally run in physical
laboratories, experiments which have well-known and widely reproduced results.
We also conducted a \emph{natural field experiment} in the sense of
the taxonomy proposed by \cite{harrison2004field}, which looked at labor
supply response to manipulations in the offered wage. Our first
experiment replicated a classic result in framing shown by
\cite{tversky1981framing}.  In accordance with numerous duplications
in the laboratory, we found that individuals are risk-averse in the
domain of gains, and risk-seeking in the domain of losses.  Subjects
were paid a fixed rate for participating.

Our second experiment had subjects play a traditional prisoner's dilemma
game.  Consistent with previous work, we found a non-zero level of
cooperation, indicating the existence of other-regarding preferences,
in which subjects do not act solely out of self-interest.  Our third
experiment had subjects play the same prisoner's dilemma game, after having been randomly assigned to read different
``priming'' passages of religious or non-religious text.  Here we
demonstrated the well-established fact that stimuli unrelated to
monetary payoffs can nonetheless affect subjects' decisions.  In both
the second and third experiments, subjects earned individualized
payments based on their choices and the choices of other workers with
whom they were randomly matched retroactively.

In our final experiment, we placed ourselves in the role of the employer.
This experimentalist-as-employer research design is perhaps the most
exciting development made possible by online labor markets.  We recruited
subjects to perform a simple transcription of a paragraph-sized
piece of text.  After performing this initial task, subjects were
offered the opportunity to perform an additional transcription task in
exchange for a randomly determined wage. As expected, we found that
workers' labor supply curves slope upwards.

\subsection{Existing research}
Several studies using online subject pools have recently appeared,
with computer scientists leading the way.  They have all used MTurk,
primarily as a way to conduct user studies and collect data suitable
for training machine learning algorithms \citep{sheng2008get,
  kittur2008crowdsourcing, sorokin51utility}.  In a paper that bridged
computer science and economics, \cite{mason2009financial} showed that,
although quality is not affected by price, output declines when wages
are lowered.  

In economics, \cite{fong2009determines} conducted an online experiment
to investigate racial bias in charitable donations, using Knowledge
Networks instead of an online labor market. Among the several
economics papers that used online labor markets, \cite{chen-wages}
measured how MTurk workers respond to wage cuts.  They found that
unexplained wage cuts decrease output, but that when the cuts are
justified to workers, the former levels of output are
maintained.\footnote{There are a number of papers that have used the
  Internet as a test bed for field experimentation, primarily as a way
  to study auctions \citep{resnick2006value,lucking2000auctions}.}  In
a separate paper using MTurk, \cite{horton2010labor} explored whether
a simple rational model can explain worker output. While they found
strong evidence that at least some workers are price-sensitive, they
also found that a non-trivial fraction are target earners, that is,
people who work to achieve certain income targets rather than
responding solely to the offered wage. \cite{chandler2010} have also
conducted a natural field experiment on MTurk.  They subtly
manipulated the meaning of the task and measured whether that meaning
affected uptake and work quality, both overall and conditional upon
a worker's home country.  Although the results are still preliminary, the
paper is inherently interesting in that it demonstrates the kinds of
experiments that would be difficult or impossible to do in other
settings.  In addition to conventional academic papers, a number of
researchers are conducting experiments on MTurk and posting results on
their blogs. Gabriele Paolacci at the University of Venice writes a
blog called
\href{http://experimentalturk.wordpress.com/}{``Experimental Turk''}
which focuses on reproducing results from experimental
psychology.\footnote{Although blogs are certainly not equivalent to
  peer-reviewed journals, they do allow academics to quickly
  communicate results and receive feedback.  For example, Rob Miller
  and Greg Little at the MIT Computer Science and Artificial
  Intelligence Laboratory (CSAIL) host a blog called
  \href{http://groups.csail.mit.edu/uid/deneme/}{``Deneme''} that
  reports the results of experiments using TurKit---a Java library
  developed by Little and others to perform iterative, complex tasks
  on MTurk \citep{little2009turkit}.}

\subsection{Replication: Framing}
Traditional economic models assume that individuals are fully rational
in making decisions---that people will always choose the option that
maximizes their utility, which is wholly-defined in terms of outcomes.
Therefore, decision-making should be consistent, and an individual
should make the same choice when faced with equivalent decision
problems.  However, as the watershed experiment of
\cite{tversky1981framing} (hereafter ``TK'') demonstrated, this is
not, in fact, the case.  TK introduced the concept of ``framing'':
that presenting two numerically equivalent
situations with different language can lead to dramatic differences in
stated preferences.  In our current experiment, we replicated the framing effect
demonstrated by TK on MTurk.\footnote{This is the second replication of
  this result on MTurk.  Gabriele Paolacci also performed this
  experiment and reported the results on his blog,
  \href{http://experimentalturk.wordpress.com/2009/11/06/asian-disease}{http://experimentalturk.wordpress.com/2009/11/06/asian-disease}.}

In TK's canonical example, subjects read one of two hypothetical
scenarios.  Half of the subjects were given the following Problem $1$:
\begin{quote}
Imagine that the United States is preparing for the outbreak of an unusual
Asian disease, which is expected to kill $600$ people.  Two alternative
programs to combat the disease have been proposed.  Assume that the
exact scientific estimates of the consequences of the programs are as
follows: If Program A is adopted, $200$ people will be saved.  If Program
B is adopted, there is $\frac{1}{3}$ probability that $600$ people will be saved
and $\frac{2}{3}$ probability that no people will be saved.

Which of the two programs would you favor?
\end{quote}

\noindent The other half were given Problem $2$ in which the setup (the first three sentences) was
identical but the programs were framed differently:
\begin{quote}
If Program A is adopted, $400$ people will die.  If Program B is adopted,
there is $\frac{1}{3}$ probability that nobody will die, and $\frac{2}{3}$
probability that $600$ people will die.
\end{quote}

The two scenarios are numerically identical, but the subjects responded
very differently.  TK found that in Problem $1$, where the scenario was
framed in terms of gains, subjects were risk-averse: $72$\% chose the
certain Program A over the risky Program B. However, in Problem $2$,
where the scenario was framed in terms of losses, $78$\% of
subjects preferred Program B.

Using these same prompts, we recruited $213$ subjects to see whether
they would reproduce this preference reversal on MTurk.  We offered a
participation fee of US\$$0.40$.  We randomly assigned subjects to a
treatment upon arrival. Consistent with TK's results, we found that
the majority of subjects preferred Program A in the domain of gains
(N=$95$: $69$\% A, $31$\% B), while the opposite was true in the
domain of losses (N=$118$: $41$\% A, $59$\% B).  The framing
significantly affected, and in fact reversed, the pattern of
preferences stated by the subjects ($\chi^2$ test, $p<0.001$).  Thus,
we successfully replicated the principle of framing on MTurk.

\subsection{Replication: Social preferences}
Another central theme in experimental economics is the existence of social
(or ``other-regarding'') preferences
\citep{andreoni1990impure,fehr1999theory}.  Countless laboratory
experiments have demonstrated that many people's behaviors are
inconsistent with caring only about their own monetary payoffs.  (For a
review, see \cite{camerer2003behavioral}.)  Here we replicated the
existence of other-regarding preferences on MTurk.  

To assess pro-social behavior on MTurk, $113$ subjects were recruited to play an
incentivized prisoner's dilemma (``PD''), the canonical game for studying
altruistic cooperation \citep{axelrod1981evolution}.  Each subject was
informed that he or she had been randomly assigned to interact with another
MTurk worker and that they both would have a choice
between two options, A or B.  In addition to a \$.20 ``show-up
fee,'' subjects were further informed of the following payoff structure 
(in units of cents): 

\begin{center} 
\begin{tabular}{c|c|c|}

    &     A (cooperate)        &     B (not)  \\
\hline A (cooperate) & 	120, 120 &  40, 160 \\
\hline B (not) & 160, 40 & 80, 80  \\
\hline
\end{tabular} 
\end{center}

Here, A represents cooperation, B represents defection, and, regardless
of the action of one's partner, playing B maximizes one's payoff. 
Rational self-interested players should therefore always
select B.  Given previous evidence from experiments in the laboratory
\citep{camerer2003behavioral}, however, we predicted that MTurk subjects
would demonstrate a level of cooperation significantly greater than $0$
in a one-shot PD.  Consistent with this prediction, we observe a level 
of cooperation significantly greater than zero ($55$\% C: sign-rank test, 
$p<0.001$) and successfully replicated with MTurk the existence of other-regarding 
preferences on MTurk.

\subsection{Replication: Priming}
Priming is a third behavioral phenomenon which we demonstrated with
MTurk.  In priming studies, stimuli unrelated to the decision task
(and which do not affect the monetary outcomes) can nonetheless
significantly alter subjects' behaviors.  Priming has attracted a
great deal of attention in psychology, and, more recently, in
experimental economics \citep{benjamin2010social}.

To employ MTurk to explore the effects of priming, we recruited 189 subjects to
play the same PD game described in the previous
section.  Subjects were randomly assigned to either the religious prime
group ($N=90$) or a neutral prime group ($N=99$).  The religious prime
group read a Christian religious passage about the importance of
charity (Mark $10$:$17$-$23$) before playing the PD.  The neutral prime group
instead read a passage of equal length describing three species of
fish before playing the PD.  Following the PD, each subject completed a
demographic questionnaire reporting age, gender, country of residence,
and religious affiliation.  The subjects also indicated whether they had ever had an
experience which convinced them of the existence of God.  Based on
previous results using implicit primes with a non-student subject pool
\citep{shariff2007god}, we hypothesized that the religious prime would
increase cooperation, but only among subjects who believed in God.

\begin{figure}
\begin{center} 
\includegraphics{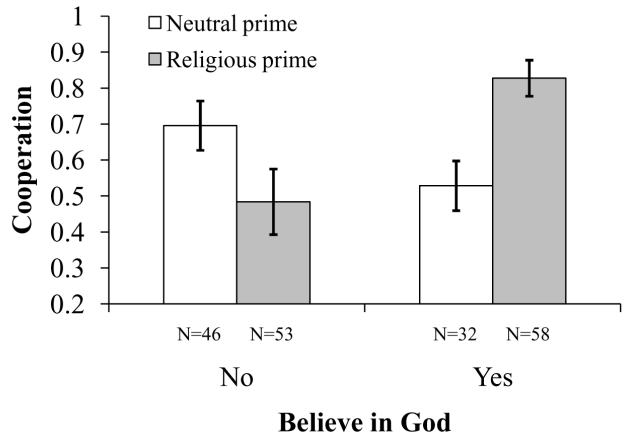}
\caption{Reading a religious passage significantly increases
  prisoner's dilemma cooperation among those who believe in God, but
  not among non-believers.\label{figX}}
\end{center} 
\end{figure}

The results are portrayed in Figure \ref{figX}.  We analyzed the data
using logistic regression with robust standard errors, with PD
decision as the dependent variable ($0$=defect, $1$=cooperate), and
prime ($0$=neutral, $1$=religion) and believer ($0$=does not believe
in God, $1$=believes in God) as independent variables, along with a
prime $\times$ believer interaction term.  We also included age, gender
($0$=female, $1$=male), country of residence ($0$=non-U.S., $1$=U.S.),
and religion ($0$=non-Christian, $1$=Christian) as control variables.
Consistent with our prediction, we found no significant main effect of
prime ($p=0.169$) or believer ($p=0.12$), but a significant positive
interaction between the two (coeff$=2.15$, $p=0.001$).  We also found a
significant main effect of gender (coeff$=0.70$, $p=0.044$), indicating that
women are more likely to cooperate, but no significant effect of age
($p=0.52$), country of residence ($p=0.657$) or religion
($p=0.54$). We demonstrated that the religious prime significantly increases
cooperation in the PD, but only among those who believe in God.  These
findings are of particular note given the mixed results of previous
studies regarding the effectiveness of implicit religious primes for
promoting cooperation \citep{benjamin2010}.  We demonstrate that the
principle of priming can be observed with MTurk and that the
effectiveness of the prime can vary systematically, depending on the
characteristics of the reader.

\subsection{Replication: Labor supply on the extensive margin}
Economic theory predicts that, under most circumstances, increasing the
price paid for labor will increase the supply of labor.\footnote{The
  exception is when the increased price changes total wealth to such
  an extent that changed tastes under the new scenario (i.e., income
  effects) might be more important than the pure substitution effect.} 
In this experiment, we exogenously manipulated the payment offered to
different workers and then observed their labor supply.  Because the
sums involved were so small, we are confident that income effects, at
least as traditionally conceived, were inconsequential in this context.  We 
found strong evidence that subjects are more likely to work when wages
are high. 

When subjects ``arrived'' at the experiment, we explained that they would
answer a series of demographic questions and then perform one
paragraph-sized text transcription for a total of 30 cents.  They were
also told that they would have the opportunity to perform another
transcription after the original transcription was completed.

\begin{figure}
  \begin{center} 
\includegraphics{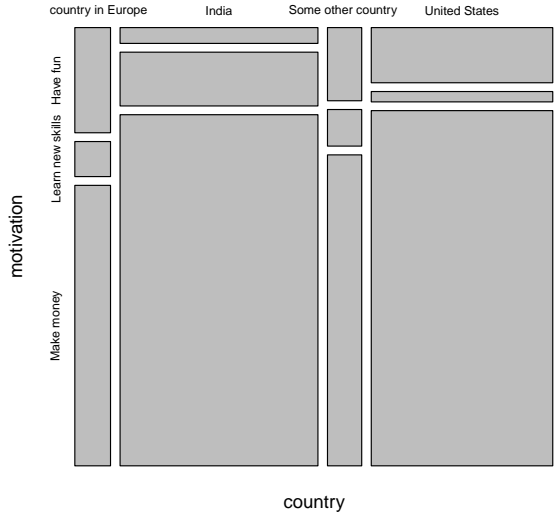}
\caption{Self-reported motivation for working on Amazon Mechanical
  Turk (row) cross-tabulated with self-reported country (column) for
  302 workers/subjects.\label{fig:mot}}
\end{center} 
\end{figure}

In addition to asking their age, gender and hours per week spent
online doing tasks for money, we asked workers to identify their home
countries and their primary reasons for participation on MTurk.
Because economic opportunities differ by country, we might expect that
motivation and behavior would also differ by country
\citep{chandler2010}.  Figure \ref{fig:mot} presents a mosaic plot
showing the cross-tabulation results.  We can see that most subjects,
regardless of nationality, claimed to be motivated primarily by money.
Among those claiming some other motivation, those from India claimed
to want to learn new skills, while those from the United States
claimed to want to have fun.

For the actual real-effort task, we asked subjects to copy verbatim
the text displayed in a scanned image into a separate text box.  The
text appeared as an image in order to prevent subjects from simply
copying and pasting the text into the text box.  The advantages of a
text transcription task are that it (a) is tedious, (b) requires
effort and attention, and (c) has a clearly defined quality
measure---namely, the number of errors made by subjects (if the true
text is known).  We have found it useful to machine-translate the text
into some language that is unlikely to be familiar, yet has no
characters unavailable on the standard keyboards. Translating
increases the error rate by ruling out the use of automated
spell-checking, and it prevents subjects from finding the true text
somewhere on the web.  For this experiment, our text passages were
paragraph-sized chunks from Adam Smith's Theory of Moral Sentiments,
machine translated into Tagalog, a language of the Philippines.

In our experiment, after completing the survey and the first task,
subjects were randomly assigned to one of four treatment groups and
offered the chance to perform another transcription for $p$ cents,
where $p$ was equal to $1$, $5$, $15$ or $25$ cents.

As expected, workers receiving higher offers were more likely to
accept.  Table \ref{tab:labor} shows that as the offered price
increased, the fraction of subjects accepting the offer rose.

The regression results are $ \bar{Y}_{i} =
\underbrace{0.0164}_{0.0024} \cdot \Delta p_i +
\underbrace{0.6051}_{0.0418}$, with $R^2= 0.13$ and
sample size $N=302$, with $\Delta p_i = p_i - 1$.  This
offsetting transformation makes the regression intercept interpretable
as the predicted mean offer uptake when $p=1$. Of course, a linear
probability model is false, as it will predict uptake rates greater
than $1$.  While we could use a general linear model, it makes more
sense to tie the problem more closely to our theoretical model of how
workers make decisions.
 
 Presumably workers' reservation wages---the minimum amount they are
 willing to accept to perform some task---have some uknown
 distribution with cumulative density function cdf $F$. Workers will
 choose to accept offers to do more work if the offered wages exceed
 their individual reservation wages. For a task taking $t$ seconds and
 paying $p_i$ cents, then $y_i = 1 \cdot \{p_i/t > \omega_i\}$,
 where $\omega_i$ is the reservation wage. If we assume that $F$ is
 the log-normal distribution, the distribution parameters that
 maximize the likelihood of observing our data are $\mu =
 0.113$ and $\sigma = 1.981$.  Given
 the average completion time on the first paragraph, the median
 reservation wage is
 US \$$0.14$/hour.

\begin{table}
\begin{center} 
  \caption{Acceptance of paragraph transcription task by offer amount \label{tab:labor}}
\begin{tabular}{lccc}
    Amount         & Offers   & Offers   & Percentage  \\
    (cents)        & Accepted & Rejected & Accepted \\
\hline  1 & 34 & 37 & 0.48 \\
5 & 57 & 18 & 0.76 \\ 
15 & 74 & 5 & 0.94 \\ 
25 & 71 & 6 & 0.92 \\
\end{tabular} 
\end{center} 
\end{table}

\subsection{Summary}
Each of these replication studies was completed on MTurk in less than
$48$ hours with little effort on our part.  The cost was also far less
than that of standard lab experiments, at an average cost of less than
\$$1$ per subject. However, even this low marginal cost vastly
understates the comparative efficiency of online experiments. We
entirely avoided both the costs associated with hiring full-time
assistants and the costs of maintaining a laboratory. We also did not
bear the high initial costs of setting up a laboratory.

Low costs would be irrelevant if the results were not informative;
yet, despite the low stakes and extreme anonymity of MTurk, the subjects'
behavior was qualitatively consistent with findings from the standard
laboratory.  Our studies demonstrate the power of MTurk to quickly and
cheaply give insights into human behavior using both traditional
laboratory-style experiments and field experiments.

\section{Internal Validity} \label{sec:intval}
It is reassuring that our experiments achieved results consistent with
those of physical laboratories, but we make an independent case for
the internal validity of online experiments.  Internal validity
requires that subjects be appropriately assigned to groups, that selective
attrition be ruled out, and that subjects be unable either to interact with or
influence one another.  These concerns could present challenges in an
online setting, where subjects can enter over time and easily drop out of an experiment.  If the level of attrition depends on
factors such as the nature of the treatment or communications from other
subjects, internal validity is jeopardized.  Fortunately, proper
measures can be taken to overcome, or at least mitigate, these challenges.

\subsection{Unique and independent observations}
MTurk and other online labor markets provide unique observations
without much effort on the part of the experimenter, because it is
difficult for workers/subjects to have multiple user accounts.  

However, if a worker had multiple accounts, in principle,
multiple plays of the experiment would be possible, which could in
turn impair the results.  At a minimum, the estimated
standard errors would be too small.  The estimated treatment effect
may be biased or even meaningless, since a repeat participant
experiences a ``treatment'' unlike any other and has no obvious
control.  In contrast, in physical laboratory experiments, multiple
plays are easily avoided.  If an experiment is conducted in a single
session, a person obviously cannot be in two places at once; if the experiment is
conducted over multiple sessions, it would be hard for a person both to use a fake ID and not be recognized.

The threat of multiple plays is surely greater online.  The anonymity
of the Internet rules out the ID approach used in the physical
laboratory, and for many experiments, subjects have an incentive to
play multiple times.\footnote{Although their per-trial payment would
  be constant, the marginal cost of completing an experiment would
  probably decrease (e.g., instructions only have to be read once).
  If skill were rewarded, then participants would benefit from
  experience.  Even leaving aside declining costs, doing an experiment
  would buy participants an option: when they found a good deal, they
  would repeat it.} Fortunately, online labor market creators have
strong financial incentives and ample resources to address the problem
of multiple accounts and its undesired offspring, multiple
plays. Online markets differ in their approaches to the problem of
multiple accounts, but all use some combination of terms-of-use
agreements and technical approaches.\footnote{ Some sites require
  workers to install custom software; presumably this software can
  detect whether multiple copies of the software are being run on the
  same computer.  Other sites charge membership fees or flat fees for
  fund transfers, both of which raise the costs of keeping multiple
  accounts.} The important question from the experimental perspective
is whether multiple accounts and plays are used sufficiently often in online experiments to threaten their internal validity?
 
While we admit that multiple accounts are both possible and more
probable online, our view is analogous to the view one must take about
possible but highly unlikely behaviors that could occur in
laboratory or field experiments.  For example, some subjects could
return with fake IDs or could explain the experiment to others who
participate on a different days or at later times.  Such behavior is
possible, yet it may actually be less likely in the online laboratory
because ($1$) as online labor markets draw on remote and diverse
pools, their members are less likely to interact or communicate and
($2$) the existence of a formal reputation system makes detectably
dishonest play costly.

Ultimately, the issue of multiple accounts is an empirical question.
In our experience to date, we have detected very low numbers of people
with multiple accounts, but we have heard anecdotes of a few
technically savvy users defeating the system.\footnote{We identify
  them by finding two accounts associated the the same IP address.} Our
view is that although multiple accounts surely exist, they are a
negligible threat in most online labor markets and are likely to
remain so. As in any well-designed regulatory scheme, the steps taken
by those who run the sites in online labor markets---and hence in
online laboratories---raise the price of prohibited behavior.  The
goal is to make undesired behavior unlikely, not impossible.
Rendering undesired behavior impossible is an unrealistic aspiration.

\subsection{Appropriate assignment}
To be certain that a treatment is having a causal effect on some
outcome, we need to assign subjects to treatment and control groups in
a way that does not depend on how they will react to a
treatment. Randomization accomplishes this goal, but even
randomization can, by chance, lead to experimental groups that differ
systematically, particularly if there are many explanatory variables. To correct for this problem after randomization we may include pre-treatment variables as regressors. Even
better, if we have complete control over assignment, as we do in the
online laboratory, we can preemptively avoid the pre-treatment
differences problem by employing a blocking design.  Such a design
identifies the key nuisance factors, creates similar groups on the
basis of that factor, and then applies the treatments within each
group.

One worker/subject characteristic is arrival time to the experiment.
We might expect early and late arrivals to an experiment to differ in
systematic ways.  For example, immediate respondents are more likely to
be heavy users compared to late respondents, and they may respond more
immediately to an experiment's advertisement.  Time zones introduce a
second difference: if an experiment is launched at $1$ a.m. PST,
experience shows that most of the early respondents (typically in the
first $12$ hours) will be from Asia and most of the late respondents
(in the next $12$ hours) will be from the Western hemisphere.

In all of our experiments, we stratify according to arrival time.  This pairs
subjects according to arrival time as closely as possible.  Thus even if arrival
behavior is related to relevant characteristics, as is likely, this
relationship will not bias the composition of the different treatment
groups.  It is important that subjects be unaware of either the
stratification or the randomization; they cannot know what treatment
is ``next,'' lest it influence their participation or behavior.

\subsection{Coping with attrition}
Subjects might drop out of the different treatments in an experiment
at different rates due to the nature of the treatments.  For instance, a video
prime that is unpleasant might prompt dropouts that a neutral video
would not. Selective attrition leads to selection bias and poses a
threat to valid inference in online experiments.  The problem is
especially acute online because subjects can potentially inspect a
treatment before deciding whether to participate.  Note that this type
of balking is different from potential subjects viewing a description
of the overall experiment and deciding not to participate.  That is
not a concern, as those people are exactly analogous to a those who view an announcement for an experiment but ignore it.

Experiments in a physical laboratory are also subject to attrition
disparities, since the participants can get up and leave if they wish.  However, it
is much easier to quit an experiment on the Internet.  Moreover, the
time investments required to ``try out'' an experiment are much lower
on the Internet.  Thus, if the burden of one treatment is greater than
that of another, MTurk subjects are more likely to drop out
selectively than their physical laboratory counterparts. The online
laboratory has two ways to deal with selective attrition. 

The solutions are to collect data on
all ``arrivals'' to an experiment and either, using Method 1, show that attrition
is consistent with a random process and assume that attrition is in
fact random, or, using Method 2, show that attrition is minimal (that is, close to
zero). It is important to note that Method 1 requires the experimenter to establish an
empirical fact which is that there is an approximate balance in collected covariates across
the groups and to establish an untestable assumption, which is that there is no
unobserved sorting that could be driving the results (that is, different
kinds of subjects are dropping out of the two groups, but by chance the
total amount of attrition is the same). When the plausibility of Method 1
is doubtful, Method 2 is superior, though it requires a more
elaborate experimental design.

The best way to eliminate attrition is to give subjects strong
incentives to continue participating in the experiment after receiving
their treatment assignments.  In the physical lab, subjects can
forfeit their show-up fee by leaving prematurely.  Online experiments
can do something similar if they employ an initial phase---identical
across treatments---that ``hooks'' subjects into the experiment and
ensures that there is no attrition after the hook phase.  For example,
all subjects might be asked to perform a rather tedious but well-paid
transcription task first, before being randomized to the relatively
easy treatment and control group tasks.  The fee for the transcription
task is forfeited if the whole experiment is not completed.

The experimenter then restricts the sample to subjects that persevere through the tedious first phase ex ante.  This increases confidence
that these subjects will remain for the following phase.  In short,
this approach has subjects invest themselves in the study in a manner that does
not differ across treatment groups.  We then raise the price of
attrition so that any differences between treatments that might drive
non-random attrition are overcome by the incentives to
comply.\footnote{Physical laboratory experiments essentially create
  the same pattern of costs, implying incentives not to quit.  Much of
  the effort for participation comes from arranging a schedule and
  traveling to the lab.}

To use this ``hook'' strategy ethically, it is important to let
subjects know initially some of the details of the experiment.  Subjects
should always know approximately what they will be doing (with
estimates of the time required) and the minimum payment they will
receive for doing it.  We have found that by providing plenty of
information at the outset and by using appropriate hooking tasks, we can
consistently drive attrition to zero.

\subsection{Stable unit treatment value assumption} \label{sec:sutva}
The stable unit treatment value assumption (SUTVA) requires that any
individual's outcome depend only upon his or her treatment assignment,
and not upon the treatment assignment or outcome of any other subject
\citep{rubin1974estimating}.  This assumption is violated when
subjects can communicate with each other about their treatments,
choices, or experiences.  

Physical laboratory experiments can avoid SUTVA problems by conducting
the full experiment in one large session and prohibiting talk during
the experiment.  In practice, experiments often need to be conducted
over several days to get a sufficient sample.  Subjects are told not
to talk to future prospective subjects, but the extent of compliance with
that request is not clear.

The SUTVA problem is both more and less challenging online than in
physical laboratories.  On the downside, the accumulation of subjects over
time is inevitable online.  The counterbalancing pluses are that the
subjects are widely geographically scattered and less likely to know
each other.  Furthermore, unlike in laboratory or field experiments, the natural
mode of conversations about goings-on in the market take place in
publicly viewable discussion forums instead of in private
encounters.

While market-specific chat rooms and forums do exist, the creators of
online labor markets make it difficult for workers to converse directly with
other workers directly in order to prevent ``disintermediation'' (moving their work offsite to escape fees).  In MTurk,
workers do not have profiles; in larger markets where workers do have
profiles, these do not contain contact information (e.g., name, email,
phone number, etc.). In MTurk, ``disintermediation'' is presumably not
as much of a concern, but Amazon is worried that if workers could
easily contact each other, they might collude in any number of ways,
such as by agreeing to approve fraudulently the work output of their
co-conspirators.  We recognize, as do the operators of online markets,
that some collusion is inevitable. Individuals in a community,
including an online community, for example, can choose to participate
in the work task and communicate outside the market.  But it seems reasonable to expect much less collusion in the online setting
than is found among the subjects in lab-based experiments, where many
repeat participants are drawn from a concentrated pool of individuals.

On the MTurk discussion boards, workers can and do highlight HITs that
they have found particularly interesting or rewarding. Sometimes they
discuss the content of the tasks.  This could skew the results of an
experiment.  Fortunately, as an experimentalist, it is easy to monitor
these boards and periodically search for mentions of relevant user
names or details from the experiment.

We have been running experiments for over a year, and occasionally
search the chat rooms for mention of our user name.  So far, it has
not appeared in any of these forums.  As a practical matter, we advise
running experiments quickly, keeping them unremarkable, and
periodically checking any associated message boards for discussions of
any experimental particulars.\footnote{It might also be possible to
  ``piggy-back'' experiments by working with existing market
  participants with established, commercial reputations---a attractive
  option suggested to us by Dana Chandler.} Warning or threatening
subjects is not recommended, as this would probably do little to deter
collusion; it might instead pique curiosity and prompt discussion.

\section{External Validity} \label{sec:extval} 
At least two factors affect the external validity of experimental
results: ($1$) its representativeness: how representative the experimental sample is of the
population of interest, and ($2$) its realism: how similar the experimental set-up
is to the real world context in which the phenomenon arises.  Any
experiment, whether in a physical laboratory or online, falls at least
a bit short on either factor.  External validity is thus a matter of
degree, not an absolute.  

On the first point, people who choose to
participate in social science experiments (or work online) represent a
small segment of the population.  Subjects in online experiments are
also highly selective compared to the U.S. population, though much
less selective than the students who make up the subjects in most
physical laboratory experiments.

In MTurk, workers from the U.S. are younger, more international, and
more likely to be female than the U.S. population, according to
self-reported demographics \citep{ipeirotis2010}. However, even if
subjects ``look like'' some population of interest in terms of observable
characteristics, the self-selection of participation is unavoidable.

On the second point, the decisions made in experiments are usually
weak imitations of the real decisions in which we are interested.
Real decisions almost always involve higher stakes, greater complexity, and greater decision time, among other differences.

The use of highly selected ``convenience samples'' is so commonplace
in experimental economics and psychology that it is rarely remarked
upon within those fields. However, social scientists from
nonexperimental fields, or at least fields with less of an
experimental tradition, often object to the use of convenience samples
and critique experiments according to both the representativeness of
the samples and the realism of the experiments. Their critiques are
worth discussing because they are central to the argument that online
experiments offer significant value.

We should first consider that the factors of representativeness and
realism are relative standards---how similar are the recruited
subjects and the experimental contexts to the real phenomena?  In any
particular experiment, theory determines similiarity.  For theories
that aspire to be general descriptions of how people make decisions or
react to stimuli, any reasonable sample of the population is
appropriate.  However, theories that cover the behavior of a selected
group (for example, young mothers) or behavior in some specialized
context (for example, after a disaster) require the sample and the
context to match for validity.  Because theories in psychology and
economics strive for generality, the behavior of a modestly broad
group of thinking human beings is usually a reliable guide.  For
example, if college students homogenize probabilities when making
decisions, it is likely that the whole population does so.

To make the point even more sharply, consider the phases of clinical drug
trials.  In the early stages, for toxicity tests, unselected healthy
humans---indeed, in the very early stages, other mammals---make fine
subjects.  In later stages, for efficacy tests, subjects must suffer
from the targeted condition.  Toxicity may differ across the
population, and it is possible that the target population for some
drug might react differently from a random sample, and yet we still
feel fairly confident using a convenience sample for toxicity but not
for efficacy.  The reason is that the questions to be
answered---respectively, ``Is it safe?'' and ``Does it
work?''---largely determine what kind of sample is needed.

\subsection{Estimates of changes are reliable:, estimates of levels are much less so} 
Quantitative research in the social sciences generally takes one of
two forms: it is either trying to estimate a level or a change.  For
``levels'' research (for example, What is the infant mortality in the United States?  Did
the economy expand last quarter?  How many people support candidate
X?), only a representative sample can guarantee a credible answer. 
For example, if we disproportionately surveyed young people, we could
not assess X's overall popularity.  For ``changes'' research (for example, Does
mercury cause autism?  Do angry individuals take more risks?  Do
wage reductions reduce output?), the critical concern is the sign of
the change's effect.  Estimating the magnitude of its effect is secondary.
Magnitudes are often assessed in specific populations once a
phenomenon has been identified.  These two kinds of empirical research
often use similar methods and even the same data sources, but one
suffers greatly when subject pools are unrepresentative, while the
other much less so.  Experiments are much more reliable for the
analysis of ``changes'' than for the assessment of ``levels.''

Laboratory investigations are particularly helpful in ``changes''
research that seeks to identify phenomena or to elucidate causal
mechanisms.  Before we even have a well-formed theory to test, we may
want to run experiments simply to collect more data on phenomena such
as the tendency to confuse nominal and real wages.  This kind of
research requires an iterative process of generating hypotheses,
testing them, examining the data and then honing or discarding
hypotheses, followed by more tests, and so on.  Because the search
space is often large, numerous cycles are needed, which gives the
online laboratory an advantage due to its low costs and speedy
accretion of subjects.\footnote{It also increases the danger of
  ``hypothesis mining''---trying out many hypotheses and reporting
  those that work, quite likely only by chance.}

\subsection{Interpreting differences between results from online and physical laboratories} 
We have found good agreement between our results and those obtained
through traditional means. Nonetheless, there are likely to be
measurable differences between results in online and physical
laboratory experiments \citep{eckel2006internet}.  How should one
interpret cross-domain differences, assuming they appear to be
systematic and reproducible?

First, systematic differences would create puzzles, which can lead to
progress. In this sense, the online laboratory can complement the
conventional laboratory and conventional laboratory---a point
emphasized by \cite{harrison2004field}. Second, it is not clear that
systematic differences should be interpreted as a mark against online
experiments. When the Internet was small and few people spent much
time online, perhaps it made sense to treat cross-medium differences
as an argument against utilizing evidence from online domains.  Today,
however, people spend a great deal of their time online.\footnote{A
  recent \emph{New York Times} article,
  \href{http://www.nytimes.com/2010/01/20/education/20wired.html}{``If
    Your Kids Are Awake, They're Probably Online,''} reported that
  American kids spend an average of seven to eight hours per day
  online.}  Internet-mediated social interactions are no longer a
strange experience familiar only to a tiny fraction of the population.

Suppose that future experiments find that subjects donate less in an
online dictator game than in person.\footnote{The dictator game is
  perhaps the simplest game design to measure social preferences:
  player 1 is given some endowment of money and must decide how much
  of that money to keep for herself and how much to give to player 2.}
This would tell us something interesting and add to our knowledge: it
would not mean that the Internet result is ``wrong,'' assuming that
the game is set up properly.  The situation is analogous to finding
cross-cultural differences in game play \citep{bohnet2008betrayal,
  gneezy2009gender}.  Such results increase our knowledge of the
social world; they are not cautionary tales about the need to restrict
experimenters to undergraduates at U.S. research universities.

\section{Experimental Designs}  \label{sec:designs}
Certain kinds of research designs work well online and are as good as,
if not better than, what can be achieved offline. The best example is
the experimenter-as-employer natural field experiments, where the
experiment in which the interaction is completely unremarkable from
the perspective of the workers, who have no idea they are involved in
an experiment. Certain kinds of surveys work well online, as we will
discuss below. Laboratory-type games are certainly possible, but there
are still a number of limitations, some of which will likely be
overcome by better software, while others are obviously intractable
due to the inherent nature of the Internet.\footnote{For example,
  recording physiological responses like eye movement, the galvanic
  skin response or blood flows to the brain are clearly not possible.}

\subsection{Experimenter-as-employer}
In many field experiments, the researcher participates in the market
as an employer, either by creating a new ``firm'' or by piggy-backing
on an existing firm and directing its policies.  Worker-subjects
perform a real-effort task, such as stuffing envelopes, picking fruit,
planting trees, etc.  The manipulation is usually of such elements as
the payment scheme, team composition, or framing. The online environment
makes it easy to conduct these kinds of experiments; subjects are
simply recruited from the market to perform a task and are randomly
assigned to groups.  The objective might be to test the role of
alternative incentive schemes or to determine how payment affects quality
of performance, both central concerns in economics.  Depending upon
institutional review board requirements, the subjects might not need to be
notified that the task is an experiment. 

For the actual onine task, it is obviously impossible to have workers
perform something physical, yet certain real-effort tasks are
well-suited to online completion.
One advantage of online work is that it generates remarkably fine
details about what workers do at any moment and time, as well as the
prevailing context when they made their chices. This kind of very
high-frequency data can yield unexpected new insights, as in
\cite{bandiera2009social} and \cite{mas2009peers}.

We have found two to be particularly useful: text transcription and
the ``dot-guessing game'' described below. Both provide good-quality
metrics and lead to heterogeneous quality, in addition to being
culturally neutral.  It is also easy to generate new instances of the
task (unlike having subjects solve puzzles). At least in the case of
text transcription, the task is very similar to the kinds of tasks
workers are frequently asked to perform on MTurk, which makes workers
less suspicious that they are involved in an experiment.

In the dot-guessing game, subjects examine an image consisting of
a very large number of dots, far too many to count feasibly, and are
asked to guess the number of dots.  This task has a clear error
metric, in that it is fully objective (that is, there is a right answer);
yet it is similar to a subjective task in that there will be
disagreement among subjects in that some subjects will presumably be better
estimators than others.  Unlike some other tasks used in judgment
experiments (for example, estimating the length of the Nile, the grain
harvest in Ukraine, etc.), subjects cannot simply look up an answer
online, which is an obvious critical difference.

A variant of the experimenter-as-employer paradigm is the
paired survey, which is used when one wants to know how some feature
of the way a question is posed affects responses. While this kind of
experiment is very simple, it can yield powerful conclusions.  For
example, some of Tversky and Kahneman's classic work, which we
replicated, used the paired-survey method to show that people viewed
objectively identical scenarios differently depending upon whether an
outcome was perceived as a gain or a loss. This simple survey design
yielded an insight that has subsequently been shown to be widespread
and important.  MTurk is ideal for this kind of research.

\subsection{Laboratory-type games online}
Some of the most successful examples of experimental social science
use simple interactive games, such as the trust game, the prisoners'
dilemma, and market games.  Subject interactions are easy to arrange
in physical laboratory experiments because all the subjects play at
once.  In online labor markets, subjects often arrive at the
experiment over the course of several hours or days, making subject
interactions difficult.  There are several solutions to the difficulty
presently available and many more under development.

When subjects are asynchronous, the widely used strategy method
\citep{selten1967strategiemethode}---players report what they would do
in various hypothetical situations---can identify outcomes in
interactive situations.  This was the method we employed when
performing our own trust-game and ultimatum-game experiments (not yet
published).  There is some evidence that subjects play ``hot'' games
(those that do not use the strategy method) differently
\citep{brandts2000hot}.  Ideally, new software developments will
allow for hot interactive games.

If the reliability of the strategy method is accepted, implementation
of online experiments is simple.  The experimenter needs only to
simulate play once all responses have been received.  This method also
has the advantage of giving more data points.  For example, in
contrast to a ``hot'' ultimatum game where we can only observe
``accept'' or ``reject'' responses from the second player, in a the
``cold'' strategy-method game, we can see the subjects' responses to
several offers because they must answer the question ``Would you
accept X?'' for several X values.

The online laboratory can allow for ``hot'' play if there are
sufficiently large numbers of respondents who can be matched up as
they arrive.  Experiments have shown that it is
possible to get MTurk workers to wait around for another
player.\footnote{See this blog post report of the experiment by Lydia
  Chilton:
  \href{http://groups.csail.mit.edu/uid/deneme/?p=483}{http://groups.csail.mit.edu/uid/deneme/?p=483}.}
This kind of approach does require that the experimenter establish
some rules for payment if a match cannot be found in a suitable amount
of time.  Another approach is to pay workers to come back to a website
at a certain time to play the game.  The great advantage of this
method is that it can be used to exactly replicate exactly the current
laboratory experience.  This method requires web-based interfaces for
games.  Work at MIT to develop ``Seaweed,'' a web-based experimental
platform, represents a strong step in this direction
\citep{chilton2009seaweed}.  Building such platforms should be a top
priority for the experimental social science community.

\section{Ethics and Community} \label{sec:ethics}
The online laboratory raises new ethical issues.  However, it also creates
opportunities for setting higher standards of evidence and for fostering
greater collaboration among social scientists, both in terms of
sharing materials and in the adversarial ``collaboration'' of replication
studies, which are now far easier to perform.

\subsection{Ethical implications of moving towards a bench science}
Online experiments can initially propel certain
subfields in the social sciences substantially toward ``bench
science.''  It is now remarkably easy to design, launch and analyze the
results of an experiment.  Conducting multiple experiments per week for
relatively small amounts of money is now feasible.  This is an
exciting and welcome development, however, in such an environment, a
researcher could turn out a stream of spuriously significant results
by burying all that are negative.  Even honest researchers can convince
themselves of the flaws in their ``pilots'' and of the legitimacy of the
subsequent experiments that happened to yield good results.

Another advantage of online experiments is that they can be run with little assistance from others.  This is coupled with the disadvantage of reducing critiques by others of procedures and results.  Since there are no
lab technicians, no subjects who can be contacted, and no logs on
university-run servers, the temptation to cheat may be high.  While
few researchers would knowingly falsify results, certain professional
norms could raise the cost of bad behavior, with the effect of both fostering honesty and
dampening skepticism.

The first norm should be machine-readable sharing of experimental
materials, as well as detailed instructions on set-up and process.
Perhaps some institution, such as a professional organization or the
National Science Foundation, could set up a library or clearinghouse
for such materials.  While most results would probably not be checked
by new experiments, requiring all experimenters to make replication
very easy would make all results ``contestable.''  This should help
make cheating an unprofitable and unpopular strategy.  Another
advantage of such a norm is that it would reduce costly duplication of
programming effort and design.  As a current example, the open-source
survey software \href{http://www.limesurvey.org}{Limesurvey} allows
researchers to export their survey designs as stand-alone files.
Researchers can simply download other people's experimental materials
and then deploy their own working versions of surveys/experiments.

There is a consensus developing in economics to make all data and code
publicly available.  To adhere to and support this norm is easy in online
contexts, but online experimenters should go a step further.  Datasets
should be publicly available in the rawest form possible (that is, in the
format in which the experimental software collected the data), as
should the associated code that turned the raw data into the data
set.\footnote{ Often it is necessary to clean this data in different
  ways, such as by dropping bad inputs, or adjusting them to reflect
  the subject's obvious intent (e.g., if a subject is asked to report
  in cents and reports $.50$, it might reasonable to infer they meant
  $50$ cents, not a half-cent). By making all trimming, dropping,
  reshaping, etc., programmatic, it is easier for other researchers to
  identify why a replication failed, or what seemingly innocuous steps
  taken by the original researcher drove the results.}  The Internet
makes such sharing a low-cost chore, since the data are invariably
generated in machine-readable form.

\subsection{Deception} 
There is a well-established ethic in experimental economics against
deceiving subjects, an ethic that yields significant positive
externalities.  Many experiments rely critically on subjects accepting
instructions from experimenters at face value.  Moreover, deception in
money-staked economics experiments could approach and even
constitute fraud.  The arguments for maintaining this ``no-deception''
policy in online experiments are even stronger.

Workers in online labor markets are a common resource shared by
researchers around the globe.  Once experiments reach a significant
scale, practicing deception in these markets could pollute this shared
resource by creating distrust.  In the online world, reputations will
be hard to build and maintain, since the experimenter's user name will be
the only thing the subject knows about the experimenter.  Of course,
economists will share the experimental space with psychologists,
sociologists and other social scientists who may not share the ``no
deception'' norm.  Probably little can be done to police other
disciplines, but economists can take steps to highlight to their subjects their
adherence to the no-deception rule and to present arguments
to others that deception is a threat to the usefulness of these
markets.  If online experiments become truly widespread, we would
expect some sites to prohibit deception, in part because their
non-experimenting employers would also be hurt by distrust.  Additional forms of
certification or enforcement processes are also likely to arise. 

\subsection{Software development} 
The most helpful development in the short term would be better (and
better-documented) tools.  There are a number of software tools under
active development that should make online experimentation far easier.
The first goal should probably be to port some variant of zTree to run
on the Internet.  The MIT initiative ``Seaweed'' is a nascent attempt
at this goal, but it needs much more development.

The research community should, as much as possible, leverage existing
open-source tools.  For example, many experiments simply require some
kind of survey tool.  The previously mentioned open-source project
``Limesurvey'' already offers an excellent interface, sophisticated
tools, and perhaps most importantly, a team of experienced and
dedicated developers and a large, non-academic user base.  Obviously
some tools, such as those for complex games, will have to be
custom-built for experiments. Advances in software are required for 
games calling for simultaneous participation by two or more subjects. Such 
software is being developed in multiple locales. 

\section{Conclusion} \label{sec:concl}
In this paper, we have argued that experiments conducted in online
labor markets can be just as valid as other kinds of experiments with the added benefit of greatly reduced cost and inconvenience. In our
replication of well-known experiments, we relied on MTurk, as MTurk is
currently the best online labor market for experimentation.  However,
as other online labor markets mature and add their own application
programming interfaces, it should be easier to conduct experiments in
other domains.  It might even be possible to recruit a panel of
subjects to participate in a series of experiments.  Many other markets have the advantage ove MTurk, whose workers remain anonymous, of offering greater ease in learning about the subjects/workers.

In this paper, we have shown that it is possible to replicate, quickly
and inexpensively, findings from traditional, physical laboratory
experiments in the online laboratory.  We have also argued that
experiments conducted in the context of online labor markets have
internal validity, and that the issue of external validity is
theory-dependent as opposed to domain-dependent.  The domains where
results will and will not have external validity are arbitrary and
require judgement by researchers.  Lastly, we have proposed a number
of new and desirable norms and practices that could enhance the
usefulness and credibility of online experimentation.  We believe that
the future is bright for the online laboratory and predict that, as
the NetLab workshop quotation in our introduction suggested, the
social sciences are indeed primed for major scientific advances.



\bibliographystyle{aer}      
\bibliography{references.bib}   

\end{document}